\documentclass[aps,prl,preprint, amsmath, amssymb, aps, groupedaddress]{revtex4-1}

\usepackage[dvipdfmx]{graphicx}
\usepackage{dcolumn}
\usepackage{bm}

\usepackage{braket}

\begin{document}


\title{Quantum Lattice Contraction Induced by Transient Raman Process}


\author{Tomobumi Mishina}
\email{mis@phys.sci.hokudai.ac.jp}
\affiliation{Department of Physics, Faculty of Science, Hokkaido University, Sapporo 060-0810, Japan}


\date{\today}

\begin{abstract}
The lattice contraction that occurs in time resolved x-ray diffraction and electron diffraction experiments 
is generally considered to be caused by photogenerated carriers.
However, quantum calculations with finite-time boundary conditions indicate that a transient Raman process
directly connects optical transitions and lattice displacements. Lattice contraction and coherent phonons
are well explained by the Raman process.    
\end{abstract}


\maketitle

Raman scattering is a basic quantum process that directly connects optical transitions and
lattice vibrations.  
Since its discovery \cite{Raman28}, it has been observed in various material
systems and now forms the basis of a standard method to evaluate the physical properties of materials.
Theoretically, Raman scattering is well explained by Fermi's golden rule (FGR) \cite{Dirac27}.

\begin{figure}[b]
\centering
\includegraphics[width=7.0cm]{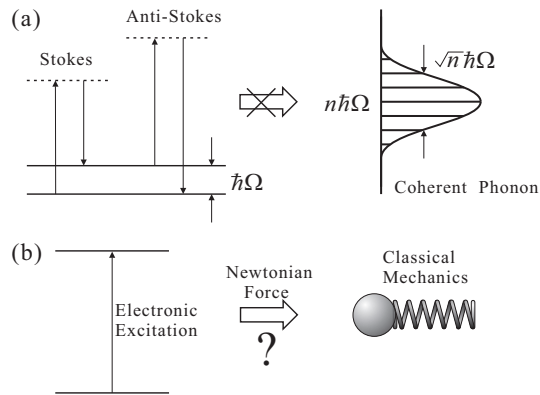}
\caption{ \label{S_Raman} 
Fermi's golden rule applied to the generation of coherent phonons.
(a) Spontaneous Raman scattering from phonon of energy $\hbar \Omega$ cannot excite the coherent phonon state 
centered at the $n$th phonon state.
(b) Other theoretical approaches consider Newtonian forces generated by electronic excitation. }
\end{figure}

Advances in ultrafast laser technology enable the generation and detection of lattice vibrations 
(so-called "coherent phonons") in the time domain.
Following the observation of the oscillation in acoustic waves \cite{Fayer78}, 
intramolecular oscillations were also observed \cite{Silvestri85}. 
The latter are explained by impulsive stimulated Raman scattering (ISRS)
in analogy with quantum beats \cite{Harde81}, which involve the interference of two optical transitions,
where each transition is explained by the FGR.   
However, for ISRS, coherent phonons are represented as a coherent state that is introduced
as the quantum counterpart of a classical oscillatory wave state \cite{Glauber63-1, Glauber63-2}.
This coherent state is a highly superposed state of energy eigenstates and cannot be excited
in the context of the FGR, as shown in Fig.~\ref{S_Raman}(a). 
Other theoretical approaches to the generation of coherent phonons consider the Newtonian forces
generated by optical transitions, as depicted in Fig.~\ref{S_Raman}(b).
To generate coherent phonons, previous works have proposed the impulsive force derived
from the dielectric energy \cite{Merlin97} and the step-like force generated by a change
in the equilibrium position of the lattice potential \cite{Zeiger92}.
The latter process is called displacive excitation of coherent phonons (DECP). 
These forces seem to explain qualitatively the generation of coherent phonon oscillations described by sine and cosine
functions, respectively. Conversely, the discovery of lattice contraction by time-resolved
electron and x-ray diffraction experiments under high-density photoexcitation of electrons \cite{Carbone08, Raman08, Runze17} 
has given rise to even greater problems. To explain this phenomenon, we consider the strong force caused by 
high-density photoexcited electrons. However, to the best of our knowledge, no clear explanation is available for the
photogenerated lattice displacement.

To address this issue, we introduce herein a quantum process to explain both coherent phonons and lattice contraction.
We show that the deep discontinuity between the Raman scattering and the generation of coherent phonons is actually
due to the temporal boundary condition in the quantum calculation.

\begin{figure}
\centering
\includegraphics[width=8.0cm]{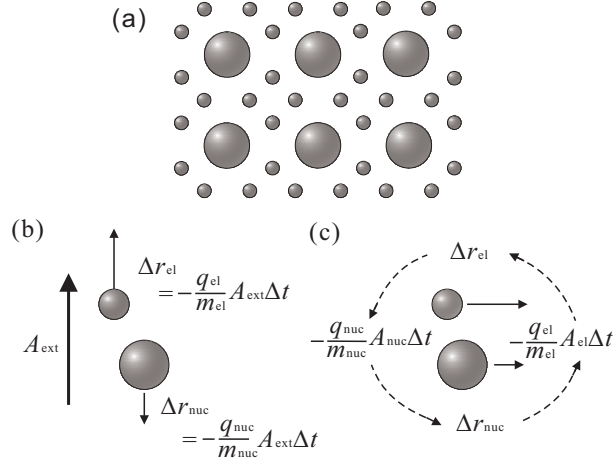}
\caption{ \label{fig:Interaction} 
Electromagnetic interaction of a crystal lattice with a laser field.
(a) Schematic drawing of lattice composed of nuclei and  electrons. 
(b) Antiphase motion of a nucleus and electron caused by direct interaction with external vector potential.
(c) Additional interaction enhances the lattice displacement.}
\end{figure}

To begin, we consider the electromagnetic interaction of a crystal lattice with a laser field.
Figure~\ref{fig:Interaction}(a) shows a schematic view of the crystal composed of nuclei and electrons.
The Hamiltonian of charged particle $i$ is 
\begin{equation}
\label{Hamiltonian}
{\cal H}^{(i)}=\frac{1}{2m_i}
\left\{ \vec{p}_i -q_i \left( \vec{A}_{\rm ext}(t)+\vec{A}_i (\vec{r}_i) \right) \right\} ^2
+q_i{\phi}_i(\vec{r}_i),
\end{equation}
where $m_i$, $q_i$, $\vec{r}_i$, and $\vec{p}_i$ are the mass, charge, position and conjugate momentum
of particle $i$, respectively. 
$\vec{A}_{\rm ext}(t)$ describes the uniform long-wavelength laser field. 
 $\phi_i(\vec{r}_i)$ and $\vec{A}_i(\vec{r}_i)$ are the scalar and vector 
electromagnetic potentials at the particle position caused by the other charged particles, respectively, and are given by 
\begin{eqnarray}
\label{Scalar_Potential}
{\phi}_i (\vec{r}_i) &=& \frac{1}{4\pi \varepsilon_0} 
\sum_{j \neq i} \frac{q _j }{|\vec{r}_i - \vec{r}_j |},\\
\label{Vector_Potential}
\vec{A}_i(\vec{r}_i)&=& \frac{1}{4\pi \varepsilon_0 c^2}
\sum_{j \neq i} \frac{q_j}{|\vec{r}_i - \vec{r}_j |}\frac{{\rm d} \vec{r}_j}{{\rm d}t},
\end{eqnarray}
where $\varepsilon_0$ and $c$ are the dielectric constant and the speed of light in vacuum, respectively.
The Hamiltonian~(\ref{Hamiltonian}) may be rewritten as
\begin{equation}
\label{Hamiltonian2}
{\cal H}^{(i)}  = {\cal H}^{(i)} _0 + {\cal H}^{(i)} _{\rm d} + {\cal H}^{(i)} _{\rm a},\\
\end{equation}
\begin{eqnarray}
\label{Sub Hamiltonian}
{\cal H}^{(i)} _0 & = & \frac{1}{2m_i}\vec{p}_i^2 
+\sum_{j \neq i} \frac{1}{4\pi \varepsilon_0} \frac{q_i q_j}{|\vec{r}_i - \vec{r}_j|},\\
\label{Direct_Interaction}
{\cal H}^{(i)} _{{\rm d}} & = & -\frac{q_i}{m_i} \vec{p}_i \cdot \vec{A}_{{\rm ext}}(t)
+\frac{q _i ^2}{2m_i}\vec{A}^2_{\rm ext}(t),
\end{eqnarray}
\begin{equation}
\label{Additional_Interaction}
{\cal H}^{(i)} _{\rm a}  = -\frac{q_i}{m_i} \left( \vec{p}_i-q_i \vec{A}_{\rm ext}(t) \right) \cdot \vec{A}_i(\vec{r}_i)
+\frac{q _i ^2}{2m_i}\vec{A}^2 _i (\vec{r}_i).
\end{equation}
Equations~(\ref{Sub Hamiltonian})-(\ref{Additional_Interaction}) correspond to
the free, the direct interaction, and the additional interaction Hamiltonians, respectively. 
The equations of motion involving  $\vec{r}_i$ and $\vec{p}_i$ for each charged particles are
\begin{equation}
\label{r_Solution}
\frac{{\rm d} \vec{r}_i}{{\rm d}t} = \frac{\vec{p}_i}{m_i}
-\frac{q_i}{m_i}\vec{A}_{\rm ext}(t)
-\frac{1}{m_i c^2} \sum _{j \neq i} 
\frac{q_i q_j}{4 \pi \varepsilon_0 |\vec{r}_i - \vec{r}_j|}\frac{{\rm d} \vec{r}_j}{{\rm d}t},
\end{equation}
\begin{equation}
\label{p_Solution}
\frac{{\rm d} \vec{p}_i}{{\rm d}t} = - \sum_{j \neq i}
\left(1 - \frac{1}{c^2}
\frac{{\rm d}\vec{r}_i }{{\rm d} t} \cdot \frac{{\rm d}\vec{r}_j }{{\rm d} t} 
\right)
\frac{\partial}{\partial \vec{r}_i}
\frac{q_i q_j}{4 \pi \varepsilon_0 |\vec{r}_i -\vec{r}_j | }. 
\end{equation}
The second term on the right-hand side of Eq.~(\ref{r_Solution}) comes from the direct interaction Hamiltonian
(\ref{Direct_Interaction}) and causes the antiphase motion of electrons and nuclei at the optical frequency,
as shown in Fig.~\ref{fig:Interaction}(b).
Conversely, the third term on the right-hand side of Eq.~(\ref{r_Solution}) comes from Eq.~(\ref{Additional_Interaction}) 
and is responsible for the positive feedback between the electron and the nucleus, as shown in Fig.~\ref{fig:Interaction}(c).
This term enhances the lattice displacement. 
The third term gives the 1/r long-range interaction network between the many charged particles,
including the core electrons and the nuclei. The problem is how to quantize this many-body interaction.

In quantum mechanics, "divergence" occurs because of the uncertainty relation.   
The FGR approximation \cite{Dirac27} is the simplest and strongest way
to exclude various processes not involved in actual observations by isolating only resonant transitions
between eigenstates over infinite time. However, the rule threatens to exclude processes
that can actually be observed.
Given the boundary conditions of finite time and space, "divergence" causes different
types of processes \cite{Stueckelberg50, Ishikawa13, Ishikawa15}. 
The quantization of the electromagnetic Hamiltonian with the boundary condition of finite space
leads to the Aharonov-Bohm effect \cite{Aharonov59}. 
The optical transition is also described by the vector potential, as in Eq.~(\ref{Direct_Interaction}). 
By using the commutator relation between the coordinates and the free Hamiltonian, 
the amplitude of the optical transition between two energy eigenstates $|E_{\rm i}> $ and $|E_{\rm f}>$
may be rewritten as
\begin{equation}
\label{FGR_approximation}
\Braket{E_{\rm f}| \frac{q}{m} \vec{p} \cdot \vec{A}_{\rm ext} |E_{\rm i}}
= \frac{\omega_{\text{fi}}}{\omega}
\Braket{E_{\rm f}| q\vec{r} \cdot \vec{E}_{\rm ext} |E_{\rm i}},
\end{equation}
where $\omega_{\rm fi}$ and $\omega$ are the transition and optical frequencies, respectively,
and $\vec E_{\rm ext}=-\partial \vec A_{\rm ext}/\partial t$ is the corresponding electric field.
In the FGR approximation, the momentum interaction is replaced with the electric-dipole interaction
because the amplitudes are identical in resonance conditions. 
However, this assumption does not hold for many nonresonant transitions, which
occur when time is finite \cite{Ishikawa17}. 
The vector potential causes the quantum motion of the nuclei 
through the momentum operators without changing the total momentum,
as shown in Fig.~\ref {fig:Interaction}(c). 

\begin{figure}
\centering
\includegraphics[width=8.0cm]{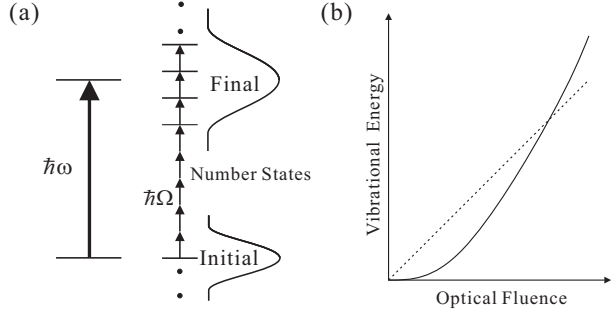}
\caption{ \label{fig:Conversion} 
(a) Optical transition and associated action of displacement operator on phonon-number states.
(b) Vibrational energy of phonon as a function of optical fluence.
}
\end{figure}

Next we derive the effective Hamiltonian for the generation of coherent phonons. 
The lattice Hamiltonian derived from Eq.~(\ref{Sub Hamiltonian}) determines the
energy eigenstates of the lattice system, including the ground state, the electronic excited states,
and the number of phonon states.
The free Hamiltonian for the phonon mode is
\begin{equation}
\label{Free_Hamiltonian}
{\hat{\cal H}}_0=\frac{\hat{P}^2}{2M}+\frac{M\Omega^2}{2}\hat{R}^2,
\end{equation}
where $\hat R$, $\hat P$, $M$, and $\Omega$ are the coordinate operator, conjugate-momentum operator,
mass, and phonon energy, respectively. $\hat R$ is the linear combination of the
displacements of nuclei from their equilibrium positions.
The interaction Hamiltonian derived from Eqs.~(\ref{Direct_Interaction}) and (\ref{Additional_Interaction}) 
gives the dielectric energy of the lattice system. 
After averaging over an optical cycle, the corresponding complex polarizability
$\hat{\chi}$ expressed in terms of phonon operators is  
\begin{equation}
\label{Taylor_Expansion2}
\hat{\chi}(\hat{R},\hat{P})= {\chi}_{0}
+ \frac{\partial \tilde{\chi}} {\partial R}\hat{R}
+ \frac{\partial \tilde{\chi}} {\partial P}\hat{P}.
\end{equation}
$\hat{\chi}$ is also a function of the laser frequency $\omega$
and contains the effects of all optical transitions.
The operator $\chi_{0}$ is responsible for the generating photoexcited charge carriers.
The interaction Hamiltonian for the phonon mode is   
\begin{equation}
\label{Interaction_Hamiltonian}
\hat{V}(t)=
-\left( \frac{\partial \tilde{\chi}_{\rm{Re}}} {\partial R} \hat{R}
+\frac{\partial \tilde{\chi}_{\rm{Re}}} {\partial P}\hat{P} \right)
\overline{E_{\rm ext}^2(t)}.
\end{equation}
If we define the phonon-annihilation operator  
\begin{equation}
\label{annihilation}
\hat{a}=\sqrt{\frac{M\Omega}{2\hbar}}\hat{R}+{\rm i}\sqrt{\frac{1}{2M\hbar\Omega}}\hat{P},
\end{equation}
then the second-quantized effective Hamiltonian is
\begin{equation}
\label{quantized_Hamiltonian}
\hat{\cal H}_{\rm eff}(t)=\frac{\hbar\Omega}{2}\left(\hat{a}\hat{a}^{\dag}+\hat{a}^{\dag}\hat{a} \right)
-\frac{1}{2}\left( \Xi \hat{a} + \Xi^{*} \hat{a}^{\dag} \right) \overline{E_{\rm ext}^2(t)},
\end{equation}
where the complex coefficient is 
\begin{equation}
\label{Interaction_coefficient}
\Xi=
\sqrt{\frac{2\hbar}{M \Omega}} \frac{\partial \tilde{\chi}_{\rm Re}}{\partial R}
-{\rm i}\sqrt{2M \hbar \Omega} \frac{\partial \tilde{\chi}_{\rm Re}}{\partial P}
=\frac{d \tilde{\chi}}{d a}.
\end{equation}
Considering $\tilde{\chi}$ to be an analytic function of $a$, Eq.~(\ref{Interaction_coefficient})
can be thought of as a complex derivative,
so the Cauchy-Riemann equations give 
\begin{equation}
\label{Cauchy_Riemann2}
\frac{\partial \tilde{\chi}_{\rm Re}}{\partial R}
=M \Omega \frac{\partial \tilde{\chi}_{\rm Im}}{\partial P},\;\;\;
\frac{\partial \tilde{\chi}_{\rm Re}}{\partial P}
=-\frac{1}{M\Omega}\frac{\partial \tilde{\chi}_{\rm Im}}{\partial R}.
\end{equation}
In the interaction picture, the time evolution of the phonon wave function
by the Schr\"{o}dinger equation gives
\begin{equation}
\label{time-evolution2}
\Psi (t) = {\cal{T}} \exp 
\left( \frac{1}{\rm{i} \hbar} 
\int_{t_0}^t 
\hat{V}_{\rm I}(t')
\overline{E^2(t')} dt'
\right) \Psi(t_0), 
\end{equation}
where $\cal T$ is the time-ordering product and
\begin{equation}
\label{Interaction Hamiltonian}
\hat{V}_{\rm I}(t)=
e^{ -\frac { \hat{\cal{H}}_0 (t-t_0)}  {\rm{i}\hbar}  } \hat{V}(t) 
e^{\frac {\hat{\cal{H}}_0 (t-t_0)} {\rm{i}\hbar} }.
\end{equation}
In the steady state, the perturbation integral averages to zero. 
If the interaction time $t-t_0$ is less than $1/\Omega$, 
the time evolution may be approximated as
\begin{equation}
\label{displacement}
\Psi (t) \approx \exp 
\left(
-\frac{\Xi {\hat a} + \Xi^{*} \hat{a}^{\dag} }{2 \rm{i} \hbar} \int_{t_0}^t \overline{E_{\rm ext}^2(t')} dt'
\right) \Psi(t_0). 
\end{equation}
The unitary operator in Eq.~(\ref{displacement}) corresponds to the displacement operator \cite{Glauber63-2}.
The coherent state is generated by applying the operator to the vacuum state, which causes
the transitions from the many-number phonon states of the initial state to those of the final state, as shown in
Fig.~\ref{fig:Conversion}(a). The amplitude of each transition is very small but their sum could be 
very large.
To evaluate the vibrational energy of the phonon mode, we consider the Heisenberg operator and its expectation value on the ground state. Because the time derivative is obtained from commutation with the total Hamiltonian, 
the rate of the vibrational energy is
\begin{equation}
\label{Vibrational_Energy}
\frac{{\rm d \Braket{ 0 | \hat{\cal H}_0 | 0 } }}{{\rm d}t}
= \Omega 
\left(
  \frac{\partial \tilde{\chi}_{\rm Im} } {\partial P} P(t) 
+ \frac{\partial \tilde{\chi}_{\rm Im}} {\partial R} R(t)
\right)
\overline{E_{\rm ext}^2(t)}.
\end{equation}
Equations~(\ref{Cauchy_Riemann2}) are used to simplify Eq.~(\ref{Vibrational_Energy}).
The equation shows that the increase in vibrational energy is
supplied by the photon absorption induced by the lattice displacement.   
The absorption is essential to generating coherent phonons  and
was already reported in early experiments in coherent phonons \cite{Fayer78}.
The amplitude of the coherent phonon is proportional to the optical fluence
and the vibrational energy of the phonon is proportional to the square of the amplitude,
so the vibrational energy is proportional to the square of the fluence,
as shown in Fig.~\ref{fig:Conversion}(b).
This dependence of the vibrational energy on the optical fluence is fully explained by the induced absorption.
Comparing Eq.~(\ref{Vibrational_Energy}) with the corresponding classical optical absorption gives the relation  
\begin{equation}
\label{Enhancement_Factor}
\Omega
\left(\frac{\partial\tilde{\chi}_{\rm Im}}{\partial P} P 
+\frac{\partial\tilde{\chi}_{\rm Im}}{\partial R }R
\right)
=\omega
\left(
\frac{\partial\chi_{\rm Im}}{\partial P} P 
+\frac{\partial\chi_{\rm Im}}{\partial R }R
\right). 
\end{equation}
The feedback process shown in Fig.~\ref{fig:Conversion}(a) between optical absorption and lattice displacement  
is neglected in the classical theory and is indicated by the factor $\omega/\Omega$.
By using the Heisenberg operator, the differential equation of the expectation value of the annihilation operator is
\begin{equation}
\label{D-annihilation}
\left( \frac{\rm d}{{\rm d}t}+i\Omega \right)a =-\frac {\Xi^*} {2{\rm i}\hbar}\overline{E_{\rm ext}^2(t)}.
\end{equation}
Integrating this equation from the equilibrium state gives
\begin{equation}
\label{I-annihilation}
a(t) =-\frac {\Xi^*} {2 {\rm i} \hbar}\int_0^t \overline{E_{\rm ext}^2(t')}e^{{\rm i} \Omega (t-t')}dt'.
\end{equation}
Applying Eqs.~(\ref{annihilation}), (\ref{Interaction_coefficient}), 
(\ref{Cauchy_Riemann2}), and  (\ref{Enhancement_Factor}) to Eq.~(\ref{I-annihilation}) and
using the impulsive condition $t<1/\Omega$, the momentum $P$ and displacement $R$  are approximated as 
\begin{equation}
\label{R&P}
\begin{split}
P(t) &\approx \frac {\partial \chi_{\rm Re}} {\partial R}
\frac{\omega}{\Omega}
\int_0^t\overline{E_{\rm ext}^2(t')}dt',\\
R(t) &\approx \frac {\partial \chi_{\rm Im}} {\partial R}
\frac{\omega}{M\Omega^2}
\int_0^t\overline{E_{\rm ext}^2(t')}dt',
\end{split}
\end{equation}
where $\partial \chi_{\rm Re}/\partial R$ and $\partial \chi_{\rm Im}/\partial R$
are the real and imaginary parts of the classical Raman tensor. 
These parts correspond to the processes formerly assigned
to ISRS \cite{Silvestri85} and DECP \cite{Zeiger92}, respectively.

\begin{figure}
\centering
\includegraphics[width=7.0cm]{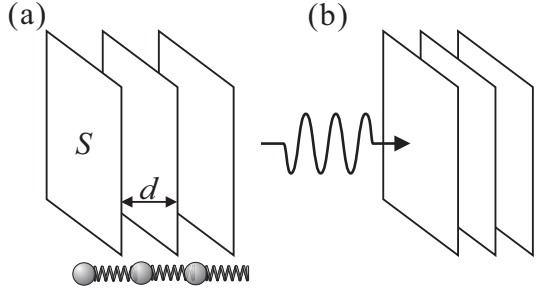}
\caption{ \label{fig:Lattice} 
(a) The ideal lattice consists of atomic layers with cross sectional area $S$ and lattice constant $d$.
(b) Photoexcitation causes the quantum lattice to contract without the photogeneration of carriers.   
}
\end{figure}

Finally, we apply the effective Hamiltonian to the ideal lattice system composed of atomic layers
with a cross-sectional area $S$ and lattice constant $d$, as depicted in Fig.~\ref{fig:Lattice}(a).
If we assume that $R$ and $P$ are the displacement of the interlayer distance and its conjugate momentum,
the Hamiltonian of the monoatomic layer is
\begin{equation}
\label{Lattice_Hamiltonian}
{\cal H}_{\rm L}
=\frac{P^2}{2\rho S d}+\frac{YSd}{2}\left( \frac{R}{d} \right)^2
-\frac{\omega Sd}{\Omega}
\left(
\frac{ \varepsilon_{\rm Re}}{\partial R}R 
+\frac{ \varepsilon_{\rm Re}}{\partial P}P  
\right)
\frac{I(t)}{c n_b},
\end{equation}
where $\rho$, $Y$, and $\varepsilon$ are the density, elastic constant,
and relative dielectric constant, respectively.
The total optical polarizability $\chi$, optical intensity $I(t)$, and 
the corresponding phonon frequency are, respectively,
\begin{equation}
\chi = Sd \varepsilon \varepsilon_0,\;\; 
I(t)=\varepsilon_0  n_{\rm b}^2 \overline {E_{\rm ext}^2(t)} c/n_{\rm b},\;\,
\Omega=\sqrt{Y/\rho}/d, 
\end{equation}
where $n_{\rm b}$ is the index of refraction of the crystal. 
We consider the change in interlayer distance in the impulsive limit $(t<1/\Omega)$.
In addition, because the acoustic velocity $v=\Omega d$,
acoustic propagation is limited by the lattice constant $d$,
so that each atomic layer is mechanically decoupled.
Replacing the parameter of Eq.~(\ref{R&P}) with the parameter of the lattice Hamiltonian Eq.~(\ref{Lattice_Hamiltonian}), 
the lattice displacement is
\begin{equation}
R \approx \frac{\partial \varepsilon_{\rm Im}} {\partial R} \frac{\omega d^2 }{Ycn_b}\int_0^t I(t')dt'.
\end{equation} 
For simplicity, the index $n_b$ of refraction is held constant. 
If the imaginary Raman tensor is negative, the equation gives a quantum lattice
contraction, as shown in Fig.~\ref{fig:Lattice}(b).
By using a lattice strain $u=-R/d$, a vacuum laser wavelength $\lambda=2 \pi c /\omega $,
and an optical fluence
\begin{equation}
F = \int_0^t I(t')dt',
\end{equation} 
the useful expression for the associated stress $\sigma$ is   
\begin{equation}
\label{Stress}
\sigma=Yu \approx \frac{1}{n_{\rm b}} 
\frac {\partial \varepsilon_{\rm Im} } {\partial u} \frac{2\pi F}{\lambda}.
\end{equation}
A fluence of $\rm{10 mJ/cm^2}$ at 800 nm corresponds to $2\pi F/\lambda$ of 0.785 GPa.
The remaining part of Eq.~(\ref{Stress}) is a dimensionless quantity of order unity.

In conclusion, we derive herein the effective Hamiltonian describing an optical-lattice interaction 
by using the complex Raman tensor and applying a finite-time boundary condition. 
The optical transition and lattice displacement are tightly linked by the transient interaction caused by 
the vector potential so that the lattice-induced absorption takes place.
The non-Newtonian momentum interaction explains
the lattice contraction and the coherent-phonon generation formerly attributed to DECP.
For further progress in this field, we encourage the lattice contraction and the induced photoabsorption to be verified qualitatively and quantitatively by experiment.
 
The author thanks Professor Kenzo Ishikawa for encouragement and stimulating discussions. 

\bibliography{QLC}

\end{document}